# Thermodiffusion mechanism for the formation of an amorphous phase during the quenching of a metal melt

A.I. Karasevskii, A.Yu. Naumuk

*G.V. Kurdyumov Instituite for Metal Physics, 36 Vernadsky Boulevard, Kiev, 03142, Ukraine*

Email: akarasevskii@gmail.com

A theoretical model of the formation of an amorphous phase during the quenching of a metallic melt is proposed. It has been shown that the appearance of a significant temperature gradient during the quenching of a metallic melt leads to thermodiffusion of defects and their outflow from the melt volume, which manifests itself in a significant reduction in the number of free spaces for the diffusion transfer of the atoms of the melt. Due to the thermodiffusion process, a significant restructuring of the microstructure of the medium occurs, which leads to a significant increase in viscosity, a decrease in the diffusion coefficient and specific volume of the substance, and changes in the mechanisms of deformation. The corresponding calculations of temperature distribution, thermodiffusion and distribution of the components of the melt have been carried out.

## 1. Introduction

The most common method for producing amorphous materials is freezing the melt. A necessary condition for the formation of an amorphous phase during quenching a metal melt is a high melt cooling rate, which is achieved, for example, by applying a thin layer of melt on a cold metal surface [1-2]. However, until now the mechanism of the effect of rapid cooling on the amorphization of the melt remains unclear.

Let us proceed from the fact that the rapid cooling of the melt requires the creation of a significant temperature gradient in the melt, and consider the effect of the temperature gradient on the state of the melt. To describe the melt, we will use a hole model of the liquid proposed by Ya.I. Frenkel [3], in which it is assumed that the liquid is a condensed medium with "holes" dissolved in it, that is, cavities into which neighboring atoms of the liquid can pass. Such a simplified model of a liquid makes it possible, at least qualitatively, to describe the whole complex of its transport and thermodynamic properties — a large self-diffusion coefficient and a low viscosity of a liquid, an increase in volume with constant interatomic distance during melting of a crystal, self-diffusion of atoms in a single-component liquid, features of compressibility of a liquid at high pressure and other [3, 4]. In the framework of the hole model, even a melt consisting of atoms of the same type can be considered as a two-component system containing atoms and holes, which, like vacancies in a crystal, diffuse into the medium, providing self-diffusion of the melt atoms.

The temperature gradient that occurs in the melt upon contact with a cold heat-conducting surface leads to thermal diffusion of the components and to a significant change in the microstructure of the medium (see, for example, [5,6]).

In this work, the thermal diffusion of holes in a liquid layer of a melt located on a cold heat-conducting surface will be investigated. It will be shown that, as a result of thermal diffusion, a flow of holes arises, directed toward the surface of the melt, which leads to the outflow of holes from the molten layer and a significant decrease in the number of free places for the diffusion motion of the atoms of the melt. Such a decrease in the number of empty places leads to a decrease in the diffusion coefficient, an increase in viscosity and a decrease in the specific volume of the medium, i.e. as a result of the thermal diffusion process, a substantial transformation of the microstructure of the medium occurs, the properties of which continuously approach the properties of the solid amorphous phase.

## 2. The temperature distribution in molten layer on the metal surface

Let on the cold, flat surface of a massive metal ($x=0$), whose temperature is $U_1$, is applied to a layer of melt with temperature $U_0$, wherein $U_0 \gg U_1$. The surface of molten metal is at $x=l$ (fig. 1).

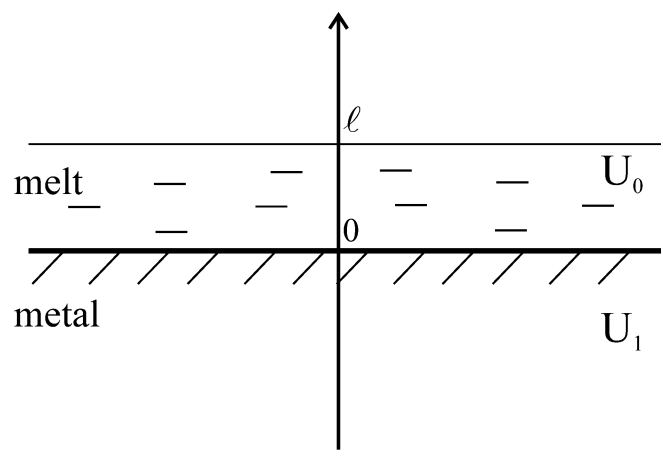

Fig. 1. A layer of metal melt on the surface of the metal

Due to the heat exchange between the melt and the metal, heat moves from the melt to the metal, which leads to a change in the temperature $u(x,t)$ of the molten metal layer. The distribution of the temperature $u(x,t)$ in the melt layer is determined by the thermal conductivity equation [7]

$$u_t = a^2 u_{xx}, \qquad (1)$$

where $u_t$ and $u_{xx}$ are the partial derivatives, respectively, in time $t$ and coordinate $x$, $a^2 = k/c\rho$, $k$ is the coefficient of thermal conductivity, $c$ - the specific heat of the melt, $\rho$ - its density. The solution of equation (1), in accordance with the statement of the problem and the choice of coordinates, must satisfy the initial condition

$$u(x,0) = U_0, \qquad (2)$$

condition of the absence of heat exchange on the free surface of the melt

$$u_x(l,t) = 0, \qquad (3)$$

and the condition of Newtonian heat exchange between a layer of a melt and a metal:

$$u_x(0,t) - h\,[u(0,t) - U_1] = 0, \qquad (4)$$

where $h$ - the heat transfer coefficient, $U_1$ - the temperature of the metal.

The solution of equation (1) in a layer thickness $l$, by the method of separation of variables [7] gives

$$u(x,t) = U_1 + \sum_{n=1}^{\infty} B_n \cos\mu_n (1 - \frac{x}{l}) e^{-\frac{t}{\tau_n}}, \qquad (5)$$

where $\tau_n = \dfrac{l^2}{\mu_n^2 a^2}$ is the characteristic time of temperature relaxation in the environment, $\mu_n = l\sqrt{\lambda_n}$, $\lambda_n$ is the eigenvalues of the boundary value problem (1) - (4).

As follows from (5), under the thermal insulation of the free surface of the melt, condition (3) is performed automatically. Taking into account the heat exchange between a layer of a melt and a metal (4) leads to the equation on $\mu_n$

$$\frac{\mu_n}{hl} = \operatorname{ctg} \mu_n. \qquad (6)$$

To determine the coefficients $B_n$, it is necessary to take into account the orthogonality of the system of eigenfunctions (5) on the interval [0, $l$] and the initial value of the temperature of the molten metal layer (2), from where

$$B_n = 4(U_0 - U_1)\frac{\sin \mu_n}{(\sin 2\mu_n + 2\mu_n)} \quad . \tag{7}$$

Taking into account (7), the expression (5) for the temperature distribution in the molten metal layer can be written as

$$u(x,t) = U_1 + 4(U_0 - U_1)\sum_{n=1}^{\infty}\frac{\sin \mu_n}{(\sin 2\mu_n + 2\mu_n)}\cos \mu_n (1-\frac{x}{l}) e^{-\frac{t}{\tau_n}} \quad . \tag{8}$$

Figure 2 shows the temperature dependence of the molten metal layer on the coordinate $x$ at different values $t/\tau$

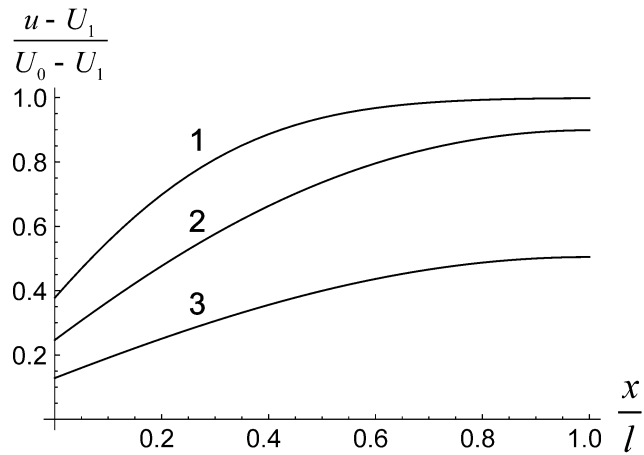

Fig.2. The given value of temperature (8) $(u(x,t)-U_1)/(U_0-U_1)$ in a melt layer at ($hl=5$, $t/\tau_n = 0.5 - 1; 1.5 - 2; 4.5 - 3$).

For further, the value of the temperature gradient is important, since, namely temperature gradient $\nabla u(x,t)$, is the driving factor in the process of thermal diffusion of the defects in the melt.

## 3. Thermodiffusion flow of defects in the melt.

The emergence of a temperature gradient in the melt at the contact of a melt with a cold metal surface initiates a thermodiffusion flux of the selected component in the melt [5-6], in our case - "holes".

$$\mathbf{J} = -(D\nabla c + D_T \nabla u), \tag{9}$$

where $D$ and $D_T$ are, respectively, the coefficients of diffusion and thermal diffusion of the defects, $c \ll 1$ is the concentration of the defects.

In the absence of convection, the distribution of the defect concentration in the melt is determined by the inhomogeneous thermodiffusion equation, which follows from the condition

$$\frac{\partial c}{\partial t} = -div\,\mathbf{J} \,, \tag{10}$$

or explicitly

$$\frac{\partial c}{\partial t} = D\nabla^2 c + f(x,t) \,, \tag{11}$$

where

$$f(x,t) = -4G_0 \sum_{n=1}^{\infty} Q_n \cos(\mu_n (1-\frac{x}{l})) e^{-\frac{t}{\tau_n}} \,, \tag{12}$$

is thermodiffusion driving force,

$$G_0 = D_T \frac{(U_0 - U_1)}{l^2}, \quad \text{and} \quad Q_n = \frac{\mu_n^2 \sin \mu_n}{(\sin 2\mu_n + 2\mu_n)}.$$

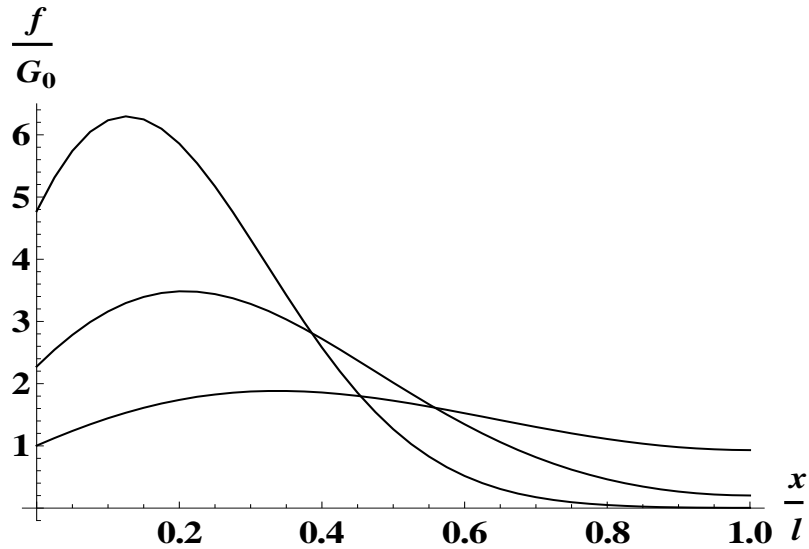

Fig. 3. Dependence $f(x,t)$ (12) on coordinates at different moments of time: $\frac{t}{\tau} = 0.25 - 1; 0.5 - 2; 1.0 - 3$.

The initial condition of equation (11) is the homogeneity of the distribution of defects in the melt:

$$c(x,0) = c_0 \,. \tag{13}$$

The first boundary condition is due to the absence of defect flux through the boundary of the metal-melt

$$c_x(0,t) = 0 \,. \tag{14}$$

The second condition is related to the diffusion output of defects through the free surface of a metal melt, which can be described as a stream of atoms through a semipermeable partition

$$c_x(l,t) = -\beta\left[c(l,t) - c_1\right], \tag{15}$$

where $\beta > 0$ is the analog of the coefficient of "porosity". In the absence of external influence $c_0$ is an equilibrium concentration of the defects in the melt, $c_1$ is the deviation of the concentration of defects from the equilibrium value in the surface area of the melt.

In the case of a temperature gradient, the distribution of defects in a melt is determined by inhomogeneous thermodiffusion equation (11), whose solution can be sought in the form of a decomposition in the Fourier series by on own functions of the equation (11) with variable coefficients

$$c(x,t) = \sum_{k=1}^{\infty} A_k(t)\cos\left(\kappa_k \frac{x}{l}\right) e^{-\frac{t}{\tau_{D,k}}} \tag{16}$$

In this case the condition (14) is automatically satisfied, and the condition (15) reduces to the equation

$$ctg\,\kappa_k = \frac{\kappa_k}{\beta l}, \tag{17}$$

where $\kappa_k = l\sqrt{\chi_k}$ - eigenvalues of the equation (11), $\tau_{D,k} = \frac{l^2}{\kappa_k^2 D}$ - characteristic time of diffusion relaxation in the medium.

Expanding also $f(x,t)$ (12) on own functions (11) we have

$$f(x,t) = -G_0 \sum_{m=1}^{\infty}\sum_{n=1}^{\infty} B(\kappa_m,\mu_n)\cos(\kappa_m \frac{x}{l}) e^{-\frac{t}{\tau_n}}, \tag{18}$$

where

$$B(\kappa_m,\mu_n) = Q_n \frac{4\kappa_m}{(2\kappa_m + \sin 2\kappa_m)} \frac{\kappa_m \sin \kappa_m - \mu_n \sin \mu_n}{\kappa_m^2 - \mu_n^2}. \tag{19}$$

Implementing the standard procedure for solving parabolic inhomogeneous equations (see, for example, [7]), we obtain:

$$c(x,t) = \sum_{k=1}^{\infty}\left\{\frac{4c_0 e^{-\frac{t}{\tau_{D,k}}}}{(2\kappa_k + \sin 2\kappa_k)}\sin\kappa_k + G_0 \sum_{n=1}^{\infty} B(\kappa_k,\mu_n)\frac{\tau_n \tau_{D,k}}{(\tau_n - \tau_{D,k})}\left(e^{-\frac{t}{\tau_{D,k}}} - e^{-\frac{t}{\tau_n}}\right)\right\}\cos\left(\kappa_k \frac{x}{l}\right) \tag{20}$$

Figure 4 shows the distribution of defect concentration in a layer of molten metal, which is established after contact with a cold surface melt.

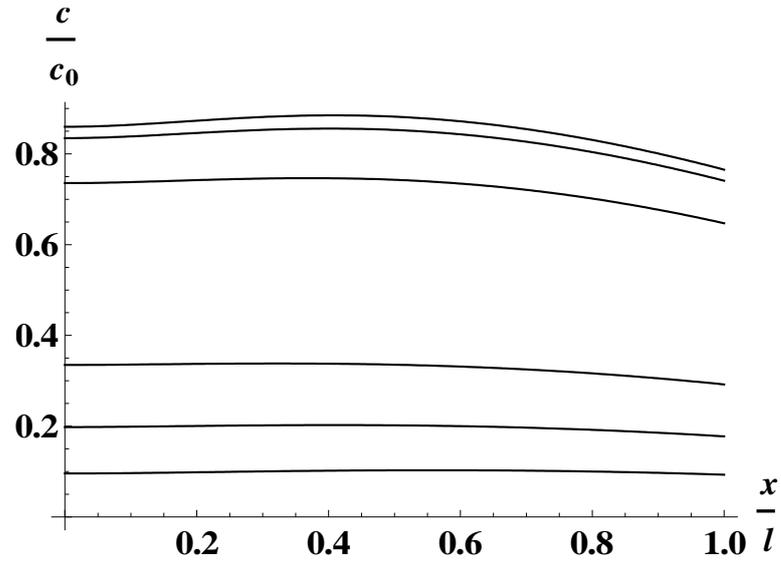

Fig.4. Distribution of the defect concentration in the melt layer: t = 0.25 tau - 1; t = 0.3 tau-2; t = 0.5 tau - 3; t = 1.5 tau - 4; t = 2 for you - 5; t = 2.5 tau - 6. Value $G_0/c_0 = 0.5$.

More informative is the time dependence of the average value of the concentration of defects in the melt layer

$$\frac{c_{avg}(t)}{c_0} = \sum_{k=1}^{\infty}\left\{\frac{4e^{-\frac{t}{\tau_{D,k}}}}{(2\kappa_k + \sin 2\kappa_k)}\sin\kappa_k + \frac{G_0}{c_0}\sum_{n=1}^{\infty}B(\kappa_k,\mu_n)\frac{\tau_n \tau_{D,k}}{(\tau_n - \tau_{D,k})}\left(e^{-\frac{t}{\tau_{D,k}}} - e^{-\frac{t}{\tau_n}}\right)\right\}\frac{\sin\kappa_k}{\kappa_k} \quad (21)$$

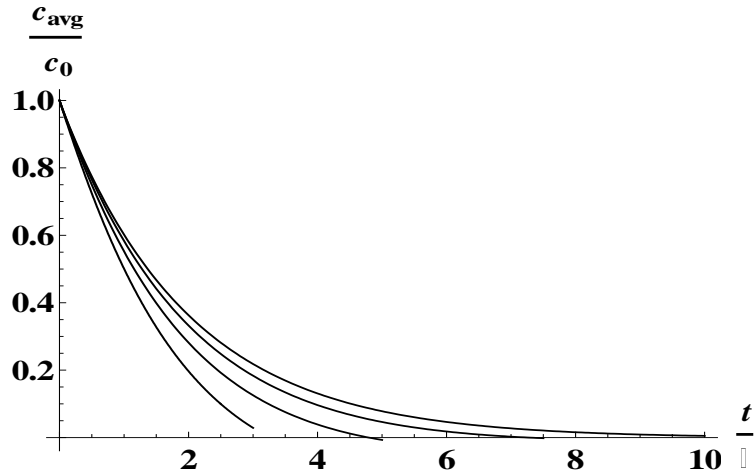

Fig. 5. Average defect concentration (28) in the molten metal layer. The value t = 0 corresponds to the beginning of the heat exchange between the melt and the metal. $G_0/c_0 = 0.5; 0.25; 0.1; 0.01$.

## 4. Results and discussion

As it follows from Figure 5, the occurrence of a temperature gradient in the melt layer results in an intense leakage of defects from the molten metal layer.

Reducing the number of structural defects, which are free places for the transition of atoms at their self-diffusion, leads to a decrease in the coefficient of self-diffusion of atoms $D_{at} = D c$ and the increase in the viscosity of the melt $\eta = \dfrac{k_B T}{6 \pi a D c}$ .

That is, in the environment there is a continuous transition from the liquid state, which is characterized by a large value of the coefficient of self-diffusion $D_{at}$ and low viscosity $\eta$ to, a solid state with practically zero value $D_{at}$ and an infinite quantity $\eta$ inherent in solids.

It should be noted that blocking the paths of diffusion (for example, oxygen) causes the corrosion resistance of the material, which is specific to amorphous materials.

As a justification of the proposed model for the formation of an amorphous state, we can refer to the results of measuring the volume reduction during the supercooling of the melt (see e.g. [1]). The transition from the crystallization temperature to the glass transition temperature of the melt is accompanied by a significant decrease in the specific volume of the medium at a practically constant interatomic distance in the melt, which can be explained only by the output of the excess volume from the melt.

It should be noted that the supercooled melt may be accompanied by the formation of quasicrystalline nanoclusters whose thermodynamic stability is due to the growth of entropy at spontaneous rotation of nanoparticles [8, 9].